\documentclass[a4paper,11pt]{article}
\usepackage[utf8]{inputenc} \usepackage[T1]{fontenc}
\usepackage{times}
\usepackage[titletoc,title]{appendix} % to get "Appendix A"
\usepackage[sf,pagestyles]{titlesec}
\titleformat{\section} {\normalfont\sffamily\bfseries} {\thesection}{1em}{}
\usepackage{amsmath}
\usepackage{amssymb}
\usepackage[dvips]{graphicx}
\usepackage{psfrag}
\usepackage[bookmarksopen=true,bookmarksnumbered=true]{hyperref}
\usepackage{calc}               % for adding lengths
\usepackage[singlelinecheck=false,labelfont={small,bf,sf}]{caption}

\usepackage{enumitem}  % allows "resume"
\setlist[enumerate]{label={\hbox to 10mm{{\small\sf\textbf{Pr.}} \small\sf\textbf{\arabic{enumi}}\hfill}}, leftmargin=12mm, itemindent=0mm, labelwidth=10mm, labelsep=2mm, itemsep=1mm}

\def\bullet{\leavevmode{\rule[0.14em]{0.24em}{0.24em}}\kern 0.33em}

\let\rhd\vartriangleright  %  
\def\csrhd{\hbox to 0 pt{\hss\hbox to 13 pt{$\rhd$\hss}}}

\let\oldsqrt\sqrt
\def\DHLhksqrt#1#2{\setbox0=\hbox{$#1\oldsqrt{#2\,}$}\dimen0=\ht0
\setbox3=\hbox{\smash{\hbox{$#1\oldsqrt{#2\,}$}}}%
\advance\dimen0-0.2\ht0
\setbox2=\hbox{\vrule width 0.4pt height\ht0 depth -\dimen0}%cs: negative depth!
\setbox4=\hbox{\smash{\hbox{\vrule width 0.4pt height\ht0 depth -\dimen0}}}%
{\box3\lower0.45pt\box4}}% cs: original from the web was 0.4pt

\normalsize

\usepackage{url}                % for url typesetting and linking, Dec 2003
\usepackage{breakurl}

\makeatletter
\def\incepsfig{\@ifnextchar[{\@incepsf}{\@incepsf[tp]}}
\def\@incepsf[#1]#2#3#4{\@ifnextchar[{\@incepsfp[#1]{#2}{#3}{#4}}{\@incepsfp[#1]{#2}{#3}{#4}[]}}
\def\@incepsfp[#1]#2#3#4[#5]{%
\begin{figure*}[#1]%
  \captionsetup{width=135mm}
  \begin{center}%
	  \scalebox{#3}{\begin{psfrags}#5{\includegraphics{#2.eps}}\end{psfrags}}%
  \end{center}%
  \caption{#4}\label{#2}
\end{figure*}}
\makeatother

\long\def\comment#1{} 
\providecommand{\email}[1]{\hbox{#1}}
\providecommand{\institute}[1]{\footnote{#1}}

\let\oldbibliography\thebibliography
\renewcommand{\thebibliography}[1]{%
  \oldbibliography{#1}%
  \setlength{\itemsep}{-2pt}}
  
\begin{document}      

\addcontentsline{toc}{section}{Comment on "Maximum force and cosmic censorship"}

\title{\Large\bfseries\sffamily Comment on "Maximum force and cosmic censorship"} % Phys. Rev. D 103, 124010 by Valerio Faraoni

\author{Christoph Schiller\ \institute{\ \ \ \ \ Motion~Mountain~Research,% ~Sperberstraße~32,
~Munich, Germany,~\email{cs@motionmountain.net} \hfill  %
        \newline\parindent 1.85em\indent\ \ \ \ \ ORCID 0000-0002-8188-6282.}}

%\date{\ } % Summer 2021

\maketitle\normalfont\normalsize\sffamily

\begin{abstract}                         % 98-90 Jan 2021, Feb 2021, Mar 2021 - F - OIWR 
\noindent\normalfont\normalsize%
Despite suggestions to the contrary, no counterargument to the principle of maximum force or to the 
equivalent principle of maximum power has yet been provided.

\bigskip % despite predicting the lack of new observable effects, 
\bigskip
\bigskip

% 83-6

\bigskip
\bigskip
\bigskip

% 83-6

% 12.15.Ff (quark and lepton masses and mixing).
% 03.67.-a (quantum information).

\end{abstract}

\newpage\normalfont\normalsize

%-----------------------------------------------------------------------------
%-----------------------------------------------------------------------------
In the introduction to a recent paper in this journal \cite{fara}, Faraoni rejects the idea of maximum force $c^4/4G$ using three arguments. 
They merit closer scrutiny.
% However, it appears that under closer scrutiny, all three arguments do not hold water.

For exploring the issue of maximum force, a useful guide is the similarity between maximum speed and maximum force.
In nature, there is an invariant \emph{local} limit on observable energy speed $c$.
Similarly, there also is an invariant \emph{local} limit on observable force $c^4/4G$.

Locality is an essential aspect in every measurement. 
An observer cannot add speeds of \emph{distant} objects and claim that their sum exceeds the local limit $c$. 
Such examples are easily found.
Instead, the speed limit $c$ states that the locally measured energy speed value, relative to the observer, never exceeds $c$.
Also, the observer performing the measurement must be physical and, e.g., cannot be superluminal.
In the same way, an observer cannot add forces of \emph{distant} mass configurations and claim that their sum exceeds the local limit $c^4/4G$.
Such examples are easily found.
Instead, the force limit $c^4/4G$ implies that the locally measured force value, relative to the observer, never exceeds $c^4/4G$.
Also, the observer performing the measurement must be physical and, e.g., cannot be a black hole, infinitesimally dense, or infinitesimally small.
These statements can also be repeated for the maximum power or luminosity $c^5/4G$.

In the paper by Faraoni, the first argument provided against maximum force is that a preprint by Jowsey and Visser \cite{jow} gives counterexamples to the value $c^4/4G$. 
However, a careful reading shows that this is not the case.
In their calculations of equatorial force values, the authors have added forces acting at different locations and found a sum larger than $c^4/4G$.
The authors have not given any example where the \emph{local} force, measured by a physical observer, is larger than $c^4/4G$.
In one additional calculation in the same work, the observer was assumed to be smaller than a black hole.
Such an unphysical setting is not a counterexample to maximum force.
This mistake is also made in a statement in \cite{s1}, where it is wrongly claimed that maximum force is valid for \emph{any} physical surface.

Since 1985, is has been stated that it is unlikely that local force values larger than $c^4/G$ or $c^4/4G$ are possible \cite{treder,sab,massa,kostro,gibbons}.
Indeed, the value $c^4/4G$ is the largest possible gravitational force between two black holes. 

Faraoni's second argument is that theories differing from general relativity do not follow maximum force \cite{nm1}.
It is not a surprise that theories that differ from general relativity do not respect the limit.
In reference \cite{me}, it was shown that maximum force $c^4/4G$ and general relativity are equivalent. 
In fact, the equivalence was the reason to call maximum force a \emph{principle} of general relativity, in the same way that maximum speed is a principle of special relativity.
The parallel between special and general relativity remains valid: 
in theories with varying speed of light, $c$ is not always a limit; in the same way, in theories with varying constants, $c^4/4G$ is not always a limit.

The third argument given by Faraoni is that counterexamples to maximum luminosity $c^5/4G$ have been published \cite{maxlum}.
However, that paper confirms that no observation with a higher luminosity value is possible in general relativity -- in contrast to other theories of gravitation -- when observers are physical.
Indeed, even the most recent observations \cite{ligo1, ligo2} fail to exceed the luminosity limit. 
The highest instantaneous luminosity observed so far is about 3\% of the maximum value.

Two remarks help clarifying the principles of relativity further.
The derivation of special relativity from the principle of a local maximum invariant speed 
-- thus from the same maximum energy speed value $c$ at every point in space and time -- 
assumes the existence of continuous and flat three-dimensional space. %
Similarly, the derivation of general relativity (in reference \cite{me}) from the principle of local maximum invariant force 
-- thus from the same maximum momentum per time value $c^4/4G$ at every point in space and time -- 
assumes the existence of continuous and smooth three-dimensional space. %

In summary, no counterexamples to maximum force or to maximum power have yet been observed or constructed, and
     no counterarguments to the principle of maximum force or to the equivalent principle of maximum power have yet arisen.
Either principle describes gravitation.

%-----------------------------------------------------------------------------
%-----------------------------------------------------------------------------
% OIWR
%

%-----------------------------------------------------------------------------

\end{document}